\documentclass[prl,twocolumn,floatfix,showpacs,superscriptaddress]{revtex4}
\usepackage{amsmath}
\usepackage{graphicx}
\begin{document}
\title{A mechanism for unipolar resistance switching in oxide non-volatile memory devices 
}
\author{M.~J.~S\'anchez}
\affiliation{Centro At\'omico Bariloche and Instituto Balseiro, (8400) San Carlos de Bariloche,
Argentina.}
\author{M.~J.~Rozenberg}
\affiliation{Laboratoire de Physique des Solides,
CNRS-UMR8502, Universit\'e Paris-Sud, Orsay 91405, France.}
\affiliation{Departamento de F\'{\i}sica, FCEN, Universidad de Buenos Aires,
Ciudad Universitaria Pab.I, (1428) Buenos Aires, Argentina.}
\author{I.~H.~Inoue}
\affiliation{Correlated Electron Research Center (CERC),
National Institute of Advanced Industrial Science
and Technology (AIST), Tsukuba 305-8562, Japan}
\date{\today}
\begin{abstract}
Building on a recently introduced model for non-volatile resistive switching,
we propose a mechanism for unipolar resistance switching 
in metal-insulator-metal sandwich structures.
The commutation from the high to low resistance state and back can be achieved 
with successive voltage sweeps of the same polarity.
Electronic correlation effects at the metal-insulator interface are
found to play a key role to produce a resistive commutation effect in qualitative
agreement with recent experimental reports on binary transition metal
oxide based sandwich structures.
\end{abstract}
\maketitle

Resistive switching phenomena controlled by external voltages
on metal-insulator-metal (MIM) sandwich structures is currently attracting a great
deal of attention due the potential applications for non-volatile memory devices.
The switching is an abrupt change between a highly resistive
state (off state) and a conductive state (on state) driven by the applied voltage.

In recent years several MIM structures where the dielectric is a 
transition metal oxide (TMO) have been reported 
\cite{beck,liu,seo}.
After electro-forming, \textit{i.e.}, the application of large voltages in order to 
produce a 
permanent change in its electric properties, those systems 
exhibit large hysteresis 
in the $I$-$V$ curves. 
In most cases
reported so far the nature of the switching is bipolar, that is, if a system 
is switched from one state to the other by a pulse of a given polarity, then a pulse of 
opposite polarity has to be applied to switch it back\cite{beck,liu}.

Surprisingly, several recent experiments on simple binary
oxides (TiO$_{2}$, NiO, CoO, etc.) demonstrated a novel 
switching effect \cite{seo, rjung, dsjeong06, dsjeong07, shima,isao}.
In these systems the switching can be achieved by successive application of voltage
pulses of the same polarity. While some qualitative ideas have been
proposed, the origin of this so called unipolar switching phenomenon remains 
largely not understood.

Recently, we have proposed a phenomenological model
that explains several
aspects of the bipolar switching phenomena \cite{prl,apl}. The goal of this letter is to
build up on that basic model and identify what additional hypothesis are
required to account for the unipolar switching effect within the same framework.
We anticipate that the key requirement is that the current injection during
the high to low resistance transition (called SET) must be of short duration (see below). 
 
The phenomenological model introduced
in Refs.\onlinecite{prl,apl} consists of an inhomogeneous dielectric medium connected to 
two electrodes (called top and bottom). The dielectric is conformed by two 
sets of small domains, each one in close physical proximity to the electrode
interfaces, and also an additional large central domain that acts as a reservoir \cite{prl}.
Carriers move from domain to domain under the action of an electric potential.
The hopping path of carriers can be associated to filamentary conduction
across the dielectric. The domains may actually correspond to grains, dislocations,
concentrated chemical inhomogeneities, such as vacancies or other type of 
defects \cite{rjung, deac, ventura06, ventura07, szot06, szot07, ddcuong, fye}. 
In the model they are
simply characterized by the number of carriers that they can contain and their
occupation at a given time. The model is further defined by the parameters
that control the probabilities for the carriers to hop from domain to domain or to
the electrodes.
An important ingredient of the model is the effect of correlations \cite{apl}. 
Since TMO are
usually prone to strong correlation effects \cite{imada}, 
we assume that under certain conditions that can be tuned by the electric pulsing,
the interface (small) domains might undergo a metal-insulator 
transition. As a result, the density of states of those domains open a charge (Mott) 
gap, and their associated hopping probabilities get greatly modified.
Since the affected domains are located at the interface, this effect has a 
significant impact on 
the conductance of the whole system.

Similarly as in Ref.\onlinecite{apl}, here we adopt a simplified version of the model
where the, in principle many, small ``top'' and ``bottom'' domains are  
represented by a single typical domain. Thus the model simply has only three domains: 
bottom, central and top.
Since the systems are usually symmetric, following Refs.\,\onlinecite{prl,apl}, 
we adopt the same hopping 
rates, $\Gamma^{ext}$, for the two interfaces [bottom(top) electrode - bottom(top) domain] 
and  the same $\Gamma^{int}$ for all the internal hopping (domain - domain). 
The model is thus defined by the following 
three coupled non-linear differential equations \cite{apl},
\begin{eqnarray} \label{eqmodel}
\frac{dn^b}{dt}
& \!\!\!\!= &\!\!\!\!\Gamma^{ext}\! f(V) \!\frac{N_{te}}{2} (1-\! \!n^b) -\! 
\Gamma^{int} f(V) n^b N_c(1-n^c),  \\
\frac{dn^c}{dt}
& \!\!\!\!= &\!\!\!\!\Gamma^{int}\! f(V) \!N_b n^b (1-\!\!n^c)- \!\!
\Gamma^{int} \!f(V) n^c N_t (1-\!n^t), \\
\frac{dn^t}{dt}
& = & \Gamma^{int} f(V)N_c n^c (1-n^t)-
\Gamma^{ext} f(V) n^t \frac{N_{be}}{2}\ \ \ . 
\end{eqnarray}

$N_\alpha$ is the total number of
states in electrode or domain $\alpha$, and $n^\alpha$ is the occupation.
$\alpha = te,be,t,c,b$ denote ``top-electrode", ``bottom-electrode'',
``top", ``central" and ``bottom". 
The voltage dependent function
$f(V)$ actually gives a time dependence to the coefficients of the differential equation
via the applied voltage protocol $V(t)$. This feature renders the system of equations very
non-linear.
A priori, the argument $V$ of these voltage dependent contribution to the
hopping probabilities should be a fraction of the externally applied voltage, and
a different one for each term in the equations above. However, for simplicity 
and to avoid a proliferation of model parameters,
we shall assume that this dependence can be absorbed in the definition of the $\Gamma$'s. 
Explicitly, we adopt $f(V)= \sinh(kV)$ \cite{simmons} where $k$ depends on various material 
parameters, and again, for the sake of simplicity, we set $k=1$. 
Positive current is defined as carriers entering the bottom domain (or leaving
the top domain). 
As in Ref.\onlinecite{apl} we assume that the electronic state of the 
domains may undergo
a metal-insulator transition.
In practice, this may occur due to injection (leakage) of
particles into (from) small domains, so that the occupation can
trigger a density driven Mott metal-insulator transition, which is
common in transition metal oxides \cite{imada}. Alternatively, it may
also occur due to a local electrochemical 
oxidation \cite{shima, szot06, szot07, ogimoto, stefanovich} that
is produced by the strong current densities during the electric 
pulsing.
The oxidation is likely to occur near the electrodes at the small 
domains \cite{isao, ybnian},
and it would locally restore the originally insulating properties of the oxide.
The phenomenon of local oxidation induced by a strong current has been observed
in Ti films \cite{cilo}. 

Upon becoming insulating, the small domains open a
charge gap $\Delta$ 
in their excitation spectrum.
We consider this effect explicitly in the model by modifying the
appropriate $f(V)$ in 
Eqs.$(1)-(3)$  to $f_{mit}(V) = e^{-\Delta/T} f(V)$, where $T$ is temperature 
(more details are given below).
Following \cite{apl} we assume that domains
are insulating when 1/2-$\delta \leq n^\alpha \leq$ 1/2+$\delta$, 
ie, when their occupation is near half-filling with about one electron per site.
We set $\delta = 0.1$.

 \begin{figure}
\centerline{\includegraphics[width=9.5 cm]{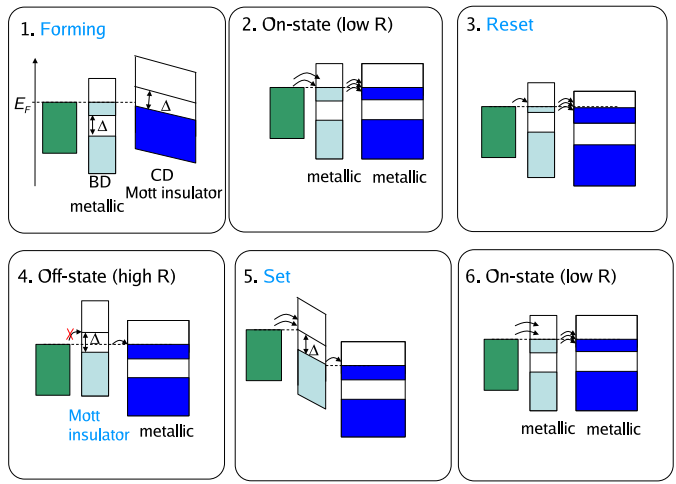}} \caption{(color on-line)
Schematic  steps of the unipolar resistive switching process. 
The number of the panels corresponds to the numbers in the 
Fig.\ref{fig2}.} \label{fig1}
\end{figure}

Within this model, a mechanism that produces the 
unipolar switching is schematically 
shown in the sequence of panels presented in Fig.\,\ref{fig1}.
Similarly as in the case described in Ref.\onlinecite{apl}, the switching
effect is mostly controlled by either one of the two small domains \cite{rjung,kmkim},
which act as ``electric faucets''\cite{isao}. 
Here we shall assume that the bottom one 
is the relevant one, while the other
remains conductive all the time. 

Panel 1 schematically indicates the initial ``electro-forming" step that is 
experimentally required. 
This process of soft breakdown is not described by our model. It involves
the metalization of the initially dielectric bulk (central domain). 
We shall simply
assume that forming sets the system in the state of panel 2, which is the
initial state of our calculation.
Thus, in panel 2, the system is in the conducting ON state (open faucet) where
the small bottom domain (BD) is metallic with an occupation above 0.6.
Upon application of a voltage pulse, 
the carrier hopping is enhanced and due to our assumption
of $\Gamma^{ext} <<\Gamma^{int}$, there is leakage from BD to the central bulk
(panel 3). 
The leakage decreases the occupation of the BD until it reaches
the threshold for the metal-insulator transition. At that point, the
BD opens a gap in its band-structure and the hopping probabilities get
reduced by a factor $e^{-\Delta/T}$ that enters in the renormalization
$f(V) \rightarrow f_{mit}(V)$, as described before.
This metal-insulator transition is thus responsible of the RESET
of the system that is now in the OFF state (panel 4).

\begin{figure}
\centerline{\includegraphics[width=9.cm]{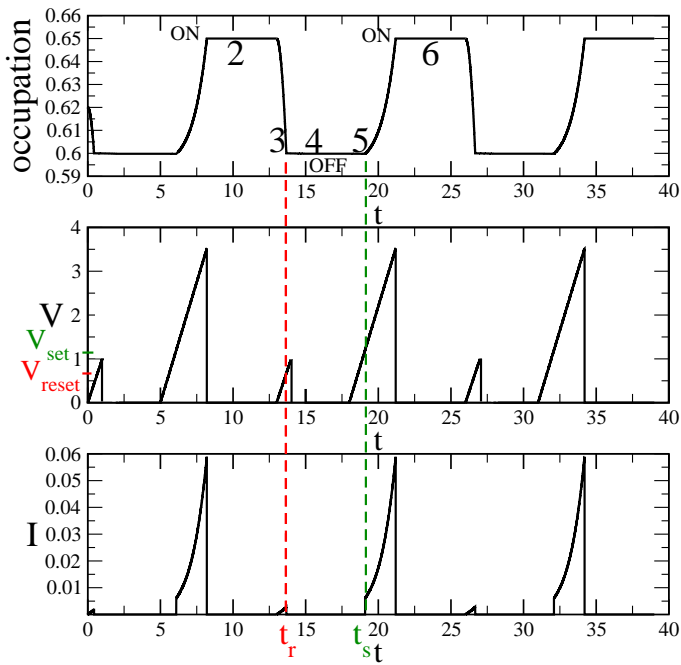}}
\caption{(color on-line) Top panel: Bottom domain occupation as a function of the time.
The numbers sequence refers to the panels of Fig. \ref{fig1}. 
The ON and OFF states are also indicated.
Middle panel: Voltage protocol as a function of time. The SET and RESET 
voltages, $V_{set}$ and $V_{reset}$,  are also shown.
Bottom panel: Current entering the BD as a function of time.
In this scale only the ON state current can be observed. The
OFF current occurs between $t_r$ and $t_s$ cannot be distinguished
from the x-axis. 
For clarity we show with dashed lines the RESET and SET transition 
times $t_r$ and $t_{s}$ for the third and fourth  
voltage pulses respectively. 
The model parameters are   $N_t=N_b=10^6$, $N_c=10^{8}$, 
$\Gamma^{int}$=10$^{-8}$, $\Gamma^{ext}$=10$^{-14}$ 
and $\Delta/T =10$.}
\label{fig2}
\end{figure}
To switch the system back to the low resistance state (SET transition)
a new voltage pulse is initiated (panel 5).
At low applied voltage a  very small current circulates since
the BD is insulating (closed faucet).
As $V$ increases, a strong electric field develops across the BD.
When the voltage at the electrode reaches a threshold value  
$V_{set}$, such that the Fermi level of the electrode aligns
with the bottom of the empty conduction band of the 
BD, a sudden injection of carriers occurs (panel 5).
As this threshold is reached, the exponential factor 
in $f_{mit}(V)$ in the first term of the rhs of Eq.(\ref{eqmodel})
disappears, 
and this $V$ dependent coefficient goes back to $f(V)$.
It is important to realize that the second term of that equation
remains unmodified as schematically shown in panel 5. 
At this point, the BD begins to increase its occupation
and eventually the BD returns to the metallic state.
This SET transition  (open faucet) is very fast; it can be associated to 
a new local electric breakdown at the BD, as it occurs at a relatively
high applied $V$. Right after the transition, 
the large applied voltage produces a large
current across the $be/b$ interface and, 
therefore, a rapid increase of the occupation of the small
domain. Thus, to prevent the BD from leaking the carriers
down to the central domain, the duration of the
applied voltage pulse should be small. It is interesting to notice that
this is reminiscent of the experimental situation, where the SET transition
has to be performed with a current control \cite{dsjeong07,hosoi}, 
which amounts to a rapid
decrease of the applied $V$ as the device becomes conducting.
After the BD returns to the metallic state (panel 6),
the factor
$f_{mit}(V)$ of the 2$^{nd}$ term of Eq.\ref{eqmodel}
also returns to $f(V)$ (open faucet) and the system
goes back to the ON state of panel 2; a new cycle may then begin.

The SET process described in detail in the previous paragraph
in which the system switches back from the
OFF state to the ON state by application of a short
voltage pulse of the {\em same polarity} as the initial reset pulse,
is the key difference between this work and that of Ref.\onlinecite{apl}.

We now substantiate the previous qualitative description by
solving the model equations following the considerations
just described.
In Fig.\ref{fig2} we show the occupation of the bottom
domain as a function of time obtained by the action of
six successive voltage pulses $V(t)$, all with positive polarity. 
In the top panel of Fig.\ref{fig2} the consecutive 
discharges and charges of the bottom domain are clearly observed. They
can be correlated with the $V(t)$ shown in the middle panel, where
the thresholds for SET and RESET are indicated.  
The current $I(t)$ entering the BD is shown in the lower panel, 
where the high current injection after the SET  
transition is clearly observed. 
On the other hand, due to the linear scale that we use, the low current
level between $t_r$ and $t_s$ cannot be distinguished.
To correlate the
successive instants of the calculation with the previous
qualitative discussion, we include in Fig.\ref{fig2}
a number
sequence that refers to the panels of Fig.\ref{fig1}.
We note that in the OFF state the occupation 
of bottom domain is smaller than 0.6, but only by an exponentially small
amount which cannot be distinguished in the linear scale used. 
There is a certain freedom
in the choice of the initial value of the occupation of the BD
[the value $n_b(t=0)$=0.62 in Fig.\ref{fig2}]. This
arbitrary initial occupation would correspond to the initial state 
after the electro-forming process which is not described by the present
model.
Nevertheless, the important observation is that after the
first sequence of RESET and SET pulses occurs, the initial
condition is lost and the occupation  
$n_b$, along with the memory operation, become cyclic.


To conclude,
the behavior of our model is found to be in good qualitative agreement with the
experimental reports of unipolar switching in transition metal 
oxide memories.
We hope that our results provide a useful guidance for further experiments needed
to improve our understanding of the basic switching mechanism in oxide non-volatile 
memories. 
The parameters of our model can be related to technologically important
features such as switching speed and non-volatility. However their explicit calculation
using ab initio methods, which would open the way for a multi-scale modeling of 
memory devices, remains a great challenge ahead.

Support from ECOS-Sud-Secyt, PICT 03-11609, PICT 03-13829 from ANPCyT, and  Fundaci\'on Antorchas 
is acknowledged.

\begin{thebibliography}{10}

  %
      
%

%
%
\bibitem{beck}%
    A. Beck {\sl et al.},
    App.\ Phys.\ Lett. {\bf 77}, 139 (2000).
%
%
\bibitem{liu}%
    S. Q. Liu, N. J. Wu, and A. Ignatiev,
    Appl.\ Phys.\ Lett. {\bf 76} 2749 (2000).
%
%
%
%
%
\bibitem{seo}%
S. Seo  {\sl et al.},
App.\ Phys.\ Lett. {\bf 85}, 5655 (2004).

\bibitem{rjung}%
R. Jung  {\sl et al.},
App.\ Phys.\ Lett. {\bf 91}, 022112  (2007). 
    
\bibitem{dsjeong06}%
D. S. Jeong,  H. Schroeder, and R. Waser
App.\ Phys.\ Lett. {\bf 89}, 082909 (2006).

\bibitem{dsjeong07}%
D. S. Jeong,  H. Schroeder and R. Waser
Electrochem.\ Solid-State Lett. {\bf 10}, G51 (2007)

\bibitem{shima}%
H. Shima  {\sl et al.},
App.\ Phys.\ Lett. {\bf 91}, 012901  (2007). 

\bibitem{isao}%
I. H. Inoue, S. Yasuda  H. Akinaga  and H. Takagi,
preprint cond-mat/0702564.

      
 \bibitem{prl}%
    M. J. Rozenberg, I. H. Inoue and M. J. S\'anchez,
    Phys.\ Rev.\ Lett. {\bf 92}, 178302 (2004).


\bibitem{apl}%
    M. J. Rozenberg, I. H. Inoue and M. J. S\'anchez,
    App.\ Phys.\ Lett. {\bf 88}, 033510 (2006).   
    

\bibitem{deac}%
A. Deac  {\sl et al.},
J. Appl.\ Phys.\ {\bf 95},6792 (2004).

\bibitem{ventura06}%
J. Ventura  {\sl et al.},
J. Alloys and Compounds {\bf 423} 181 (2006).

\bibitem{ventura07}%
J. Ventura, Z. Zhang, Y. Liu, J. B. Sousa, and P. P. Freitas,
J. Phys.: Condens. Matter 19 176207 (2007),

\bibitem{szot06}%
K. Szot, W. Speier, G. Bihlmayer, and R. Waser,
Nature Materials {\bf 5}, 312 (2006).

\bibitem{szot07}%
K. Szot, R. Dittmann, W. Speier, and R. Waser,
Phys.\ Stat.\ Sol.\ (RRL) {\bf 1}, R86 (2007).

\bibitem{ddcuong}%
D. D. Cuong  {\sl et al.},
Phys.\ Rev.\ Lett. {\bf 98}, 115503 (2007).

\bibitem{fye}%
F. Ye  {\sl et al.},
J. Appl.\ Phys.\ {\bf 101}, 113528 (2007).

\bibitem{imada}%
 M. Imada, A. Fujimori and Y. Tokura,
    Rev.\ Mod.\ Phys. {\bf 70}, 1039 (1998).
%
%
%
%
\bibitem{simmons}%
    J. G. Simmons and R. R. Verderber,
    Proc.\ Roy.\ Soc.\ A {\bf 301}, 77 (1967).
%
%
%
%
%
%
%
%
%

\bibitem{ogimoto}%
Y. Ogimoto, Y. Tamai, M. Kawasaki, and Y. Tokura,
App.\ Phys.\ Lett. {\bf 90}, 143515 (2007).

\bibitem{stefanovich}%
G. B. Stefanovich, C.-R. Cho, E.-H. Lee, I. K. Yoo,
J. Non-Crystal.\ Solids {\bf 353} 956 (2007).
 
\bibitem{ybnian}%
Y. B. Nian, J. Strozier, N. J. Wu, X. Chen, and A. Ignatiev
Phys.\ Rev.\ Lett. {\bf 98}, 146403 (2007).
 
\bibitem{cilo}
T. Schmidt, R. Martel, R. L. Sandstrom, and Ph. Avouris,  
Appl. Phys. Lett. {\bf 73}, 2173-2175 (1998).

\bibitem{kmkim}%
K. M. Kim, B. J. Choi, Y. C. Shin, S. Choi, and C. S. Hwang 
App.\ Phys.\ Lett. {\bf 91}, 012907 (2007).

\bibitem{hosoi}%
Y. Hosoi  {\sl et al.},
\textit{Tech.\ Dig.\ ---Int.\ Electron Devices Meet.} (IEEE, New York, 2006), p.793.


%
\end{thebibliography}
\end{document}